\documentclass[12pt]{article}
\usepackage{cite, ulem}
\usepackage{amsmath}
\usepackage{amssymb}
\usepackage{braket}
\usepackage{footnote}
\makesavenoteenv{minipage}

\newcommand{\mc}{\mathcal}
\newcommand{\p}{$\mathcal{PT}$ \ }

\renewcommand{\title}[1]{%
    \bigskip%
    \begin{center}%
    \Large\bf #1%
    \end{center}%
    \vskip .2in}

\renewcommand{\author}[1]{%
    {\begin{center}
    #1
    \end{center}}}
\newcommand{\address}[1]{\vspace{-1.7em}\vspace{0pt}
    {\begin{center}
    \it #1
    \end{center}}}

\begin{document}

\begin{titlepage}
 \title{Removal of instabilities of the higher derivative theories in the light of antilinearity}
\author
{
Biswajit Paul  $\,^{\rm a,b}$,
Himangshu Dhar $\,^{\rm a,c}$,

Biswajit Saha    $\,^{\rm a, d}$}
\address{$^{\rm a}$ National Institute of Technology Agartala \\
 Jirania, Tripura -799 055, India
\footnote{
{$^{\rm b}$\tt biswajit.thep@gmail.com,}
{$^{\rm b}$\tt drbiswajit.phy@nita.ac.in,}
{$^{\rm c}$\tt himangshu171@gmail.com,}

{$^{\rm e}$\tt biswajit.physics@gmail.com}}
}
\begin{abstract}
Theories with higher derivatives involve linear instabilities in the Hamiltonian commonly known as Ostrogradski ghosts and can be viewed as a very serious problem during quantization. To cure {this} , we have considered the properties of antilinearty that can be found inherently in the non-Hermitian Hamiltonians. Owing to the existence of antilinearity, we can construct an operator, called the $V$-operator, which acts as an intertwining operator between the Hamiltonian and its hermitian conjugate. We have used this $V$-operator to remove the linear momenta term from the higher derivative Hamiltonian by making it non-Hermitian in the first place via an isospectral similarity transformation. The final form of the Hamiltonian is free from the Ostrogradski ghosts under some restriction on the mass term.  
 
\end{abstract}

\end{titlepage} 
	\section{Introduction}

From usual quantum mechanics, it is well known that the Hamiltonian must be Hermitian, i.e. $H=H^\dagger$, in order to obtain real energy spectrum.  Very recently Bender et al proposed that parity ($\mc{P}$) and time ($\mc{T}$) symmetry can serve as a better condition for obtaining the real energies of the system as it includes  non-Herimitian\cite{bender2002, bender2004, bender2005, Bagchi:2000rr} Hamiltonians too. $\mc{PT}$-symmetry is actually a  physical condition obeyed by almost every phenomenon. In this regard, we may consider a non-Hermitian system {by}   employing the condtions of $\mc{PT}-$symmetry one can obtain real energy spectrum \cite{bender1998}. 
While working in the same course of considerations of the physical interpretations of the non-Hermitian Hamiltonians, one may consider  the obvious nature of these non-Hermitian Hamiltonians i.e. the antilinear property.  Evidently, the Hamiltonian that is non-Hermitian can  be written in an antilinear form and the corresponding antilinearity operator can be obtained \cite{Mannheim:2017apd,Mannheim:2015hto}. The reality of the energy eigenvalues and unitarity of the system was {found}  to be more subtle in this case. 
 
A common source of these non-Hermitian Hamiltonians is the  Higher Derivative(HD) theories. By higher derivative, we refer here to the theories that are having time derivatives of the fields more than two in the Lagrangian. The higher derivative terms are added to the Lagrangian as a correction term and these may lead to {avoid}  the ultraviolet divergences appearing in the theory.  Due to this nature,  the Higher derivative theories are actively under consideration in various branches of physics e.g. string theory \cite{stelle, neupane}, cosmology \cite{nojiri4,Gama}, general relativity \cite{soti,gullu, woodard} etc. In the HD theories, the Hamiltonian actually consists of the momenta of the higher derivative fields multiplied by the momenta of other fields and not corresponding to its own.  These linear momenta, while quantization, can lead to instabilities as the spectrum become infinite. This is a very classic problem in physics  and the corresponding field is known to be  `ghost state' or Ostrogradskian instability.

To remove these ghost states, there have been different attempts made by many authors. Like in   \cite{chen, Jimenez:2020dpn}, the authors  have tried to remove the ghost states by incorporating new constraints in the phase space which is applicable only if the phase space is reduced. Very recently, the authors in \cite{Aoki2020} considered the inclusion of velocity dependent constraints to remove the Ostrogradskian instability. The Ostrogradski ghosts can also be removed by the introduction of new variables which can be  \textbf{obtained} by a combination of primary and secondary constraints\cite{Klein:2016aiq,Paul:2017mcw}. By considering degenerate Lagrangian which has a non-invertible kinetic matrix, the theory can be made ghost-free as shown in \cite{Langlois:2015cwa}. In the case of the analytic mechanics, the ghosts appeared as usual \cite{Motohashi:2014opa} and they were removed {by}  different degeneracy conditions \cite{Motohashi:2018pxg}.  It is thus seen that there are multiple attempts to address the issue of Ostrogradski instability with limited applicability.

In this paper, we shall consider the properties of antilinearity in order to remove the instabilities from the theory. Due to antilinearty, the Hamiltonian of these kinds of models can be put into a form so that they remain invariant under a suitable choice of similarity transformation. It was later shown that the similarity transformation mentioned here, in general, can be found in any real Hamiltonian and the corresponding operator can be  identified\cite{Mannheim:2009zj, Mannheim:2015hto}. This, however, is not the case when one has a non-Hermitian  Hamiltonian $H \ne H^\dagger$. As the Ostrogradski ghosts are inherent in any HD theory, we can look into the matter by transforming it into a complex form and analyzing them using the properties due to antilinearity.  As an example,  we have considered is the Gallilean invariant Chern -Simon's model. This is a toy model which manifestly involves the characteristics of Chern-Simon's model with a mass term. Lukiersky et al. have shown that the model can be quantized in the non-commutative plane \cite{lukiersky}. This  is a interesting model as one can see that the model has been used in different fields like  quantum gravity  \cite{Papageorgiou:2009zc}, Newton-Hooke symmetry \cite{Alvarez:2007fw, Horvathy:2004fw}, anyons \cite{Horvathy:2004fw} etc. 


The plan of the paper is as follows. In Sec II we consider a very brief discussion on higher derivative models and how one can construct the non-Hermitian Hamiltonian as a necessary transformation. Sec III deals with the properties of antilinear Hamiltonians. Here, we also show how HD theories have an inherent property of anti-linearity. In Sec IV we have considered an example, the Gallilean invariant Chern-Simon's model, to illustrate the above discussion of the removal of the Ghosts. Finally, we conclude in {Sec}  V.

	\section{Higher derivative models}
    
    We may write a general HD lagrangian in the form of
    $\mc{L}(q, \dot{q}, \ddot{q} ... q^{(n)})$ which represents a theory with $n^{th}$ order derivatives in time. Now we can write the Lagrangian with respect to some new variable $Q$ defined as
    \begin{equation}
    Q_0 = q,
    Q_1 = \dot{q} ,  
    Q_2 = \ddot{q}, 
     ... 
    Q_n = \frac{d^n}{dt^n}q . 
    \end{equation}
This  redefinition of the space variables has expanded the configuration space thereby increasing the number of degrees of freedom. In the newly defined configuration space, we can see the  new constraints which are given by
$\Phi _ n = Q_n - \dot{Q}_{n-1}$. Hence the Lagrangian can be redefined by incorporating these constraints in the Lagrangian via Lagrange's undetermined multipliers as 
   \begin{eqnarray}
  \mc{L}' =  L(Q, Q_1, Q_2 ... Q_n) + \sum_{i=1}^{n} \lambda_i  \Phi_i.
   \end{eqnarray}
   The above Lagrangian is in the first order form and apparently free from the Higher order derivatives. Thus, when written in this form we can easily find the appropriate phase space ($Q_i,P_i$). Unlike the Ostrogradski way \cite{ostro} of defining the momenta, in the first order formalism, the momenta is defined in the usual way as $P_i = \frac{\partial  \mc{L}'}{\partial \dot{Q}_i}$. Immediately we write the canonical Hamiltonian which is given by 
   \begin{equation}
   \mc{H}_{can} =\sum_{i=0}^{n} P_i \dot{Q}_i - \mc{L}'.
   \label{canham1}
   \end{equation} 
   The definition of canonical momenta contains variables from the phase space which, in this case, may contain the non-invertible momenta. Usually, the higher derivative theories contain at least one constraint resulting due to the definition of momenta of higher derivative fields irrespective of the Ostrogradskian or the first order formalism.  These, in the final form of the canonical Hamiltonian, always appear in terms involving the product of fields in first order and linearly coupled canonical momenta. Interestingly, this momenta is not the {one} corresponding  to the field to which it is linearly coupled with. In the corresponding quantum picture, the linear momenta terms give rise to infinities, and hence a series of unstable states known as ghost states are being created. Next, we write the canonical momenta (\ref{canham1}) in a generic form showing the involvement of the linear momenta term as 
   \begin{equation}
   \mc{H}_{can} = \sum_{i=1}^{n} P_i Q_{i-1} + \bar{\mc{H}}. \label{canham2}
   \end{equation}
   Where, $\bar{\mc{H}}$ contains terms that do not include any linear momentum. One should remember that the term $\bar{\mc{H}}$ here is not general, as one can always find the remaining part in the canonical Hamiltonian as a  term arising due to the higher derivative nature of the theory.  Collecting all the primary constraints,  the total Hamiltonian of this system can be written as 
   \begin{equation}
   H_{T} = \mc{H}_{can} + \sum_{i=1}^{m} \Lambda_i \Phi_{i}.
   \end{equation}
   Where, $m$ represents the number of primary constraints arising in the theory. {Till now our discussions were purely based on the classical picture. In the next section, we shall consider the corresponding quantum version of the system. For the transition from classical to quantum, we shall replace the variables with their quantum counterpart and also replace all the Poisson brackets with corresponding commutators.}
   
   \section{Antilinearity in general quantum theories}
      
   The wave function $\ket{\Psi(r,t)}$ when acted upon by the total  Hamiltonian gives the energy of the system as
   \begin{equation}
   \hat{H}_{T} \ket{\Psi(r,t)} = E \ket{\Psi(r,t)}. 
   \end{equation}
   
   If we replace $t$ by $-t$ and considering an antilinear operator $A$ we can rewrite the above expression as
   \begin{eqnarray}
   A\hat{H}_{T}  A^{-1} A \ket{\Psi(r,-t)} = E^* A\ket{\Psi(r,-t)}.
   \end{eqnarray}
  Thus, for the state $\ket{\Psi(r, -t)} = A\ket{\Psi(r,t)}$, we can consider that  under the action of  operator $A$, the  Hamiltonian is 
   \begin{equation}
   \hat{\tilde{H}} =  A\hat{H}_{T}  A^{-1}. \label{8}
   \end{equation}
   \textbf{Evidently, this similarity transformed Hamiltonian also has the energy eigen value $E^*$ corresponding to the eigen state $\ket{\Psi(r, -t)}$:}
  
   \begin{equation}
   \hat{\tilde{H}} \ket{\Psi(r, -t)} = E^* \ket{\Psi(r, -t)}. 
   \end{equation}
   Energy eigenvalues can be real or imaginary if the Hamiltonian $\hat{H}_T$ of the system poses antilinear symmetry as shown in  \cite{Mannheim:2015hto, Mannheim:2009zj}. On the other hand, the reality of energy eigenvalues was also considered by Wigner in \cite{wigner}, which says about the necessity of the time reversal symmetry of the system. Thus, for the non-Hermitian Hamiltonians, it is possible to possess real energy spectra in the presence of a time reversal symmetry.
   
   In order to find an appropriate form of the $\hat{A}$ operator discussed here, we can look into the theories having $\mc{PT}$ symmetries. In $\mc{PT}$ symmetries, the theory may not be  invariant under  individual transformations of  $\mc{P}$ (which effects as $\hat{x} \rightarrow -\hat{x}, \hat{p} \rightarrow -\hat{p}$ ) and $\mc{T}$  ( which effects as $\hat{x} \rightarrow \hat{x}, \hat{p} \rightarrow -\hat{p}, i \rightarrow -i$), but,  under the collective effect of $\mc{PT}$, the theory should remain invariant. Apart from the $\mc{PT}$ operator, Bender et. al. also pointed out another operator called the $\mc{C}$ operator \cite{Bender:2003fi} that can be used to remove the ghost states as shown in \cite{bender1, BHMB}. The existence and completeness of the theory demands that the  $\mc{C}-$operator should be governed by the three conditions, which are
   \begin{equation}
   \mc{C}^2 = \mc{I}, \ \ \ \ \ \ \ \ \ [ \mc{C}, \hat{H} ] =0, \ \ \ \ \ \ \ \ \ [ \mc{C}, \mc{PT} ] =0. \label{c_properties}
   \end{equation}
   The existence of the $\mc{C}$ operator also guarantees that the unitarity of the theory will be preserved. 

   In case the Hamiltonian is not Hermtian, it should obey the following relation
   
   \begin{equation}
   \hat{V} \hat{H} \hat{V}^{-1} = \hat{H}^\dagger, \label{11}
   \end{equation}
   
   which is obtained by  defining the new operator\cite{Mannheim:2009zj}
   \begin{equation}
   \hat{V} = \hat{A}^\dagger \hat{A}. \label{12}
   \end{equation}
 
{These two properties in (\ref{11},\ref{12}) also confirm that the transformation under these operators is unitarily equivalent. The relations (\ref{8}) and (\ref{11}) both refer to connecting two different transformations of the Hamiltonian and for this reason, they are called intertwining operators.}
In the present case of higher derivative systems, the Canonical Hamiltonian $\hat{H}_{can}$  has a non-Hermitian form and the above equation will serve  a very important role in eliminating the ghost states. For that, we will consider the similarity transformations of the fields i.e. $\hat{V} Q \hat{V}^{-1}$ and a proper choice of the $\hat{V}$ will give us the Hamiltonian which is  free from the Ostrogradski instability appearing in (\ref{canham2}).
	
Below we note down  {the}  properties arising due to the antilinearity present in the theory:
\begin{itemize}
	\item If the Hamiltonian has antilinear symmetry then the energies may be real or complex. The complex energy eigen values must appear in  conjugate pairs,
	\item If the parity operator obeys $\mc{P}\hat{H}\mc{P}^{-1} = \hat{H}^{\dagger}$ then we can write  $[\mc{P}\hat{V},\hat{H}]=0$,
	\item One can relate  the $\mc{C}-$operator as $\mc{C} = \mc{P}\hat{V}$ which should obey the relations of $\mc{C}$  mentioned  in (\ref{c_properties}) ,
	\item The unitary evolution of the system is governed by the condition $\bra{\Psi_i}\hat{V}\ket{\Psi_j}= \delta_{ij}$ and we may treat this as the new definition of the inner product. This will arise only if the Hamiltonian is not Hermitian and consequently the basis states will not be orthonormal in the Dirac sense. \cite{mannheim2018}
	\item  The completeness relation in this case becomes $\Sigma |{\Psi_i}\rangle \langle {\Psi_j}|\hat{V} = \Sigma \hat{V}^\dagger |{\Psi_j}\rangle \langle {\Psi_i}| = \mc{I} .$
	
\end{itemize}	
 \subsection{Ghost states and antilinearity}
 
  {The Ostrogradski ghost problem appears due to the reason that there is no \textbf{lower} bound to the potential and consequently we cannot define a true minimum which we call the vacuum. This concept of vacuum is totally a quantum concept and  {does}  not exist {at}  the classical level. As soon as one puts $\hbar \rightarrow 0$, the quantum phenomenon lose their existence. Therefore, the quantum concepts are very much required to discuss the Ostrogradski ghost problem.}
  
The phase space of HD theories is spanned by the momenta of the usual as well as the HD fields. In this subsection, we show the connection between  HD theories and antilinearity. The existence of the higher derivative linear momenta terms make the Schrodinger equation complex and therefore, to solve them,  one requires complex planes. However, while quantizing,  as the HD momenta span the entire imaginary plane, it is seen that the solution does not give an well behaved wave function \cite{bender1998}. {For the well behaved wave function, one of the condition says that it must vanish at infinity. In the complex plane, the wavefunctions must also vanish asymptotically along lines that are centered about the positive-real and negative-real axes. In complex geometry, these lines are known as Stokes wedges. The angular opening of the Stokes wedges depends upon the type of eigenfunctions. To disallow the solutions spanning the whole phase space, we can restrict them within  Stoke wedges of $90^\circ$.}   For avoiding the imaginary axis, we can write the Hamiltonian (\ref{canham2}) after an isospectral similarity transformation, of the usual field (not the HD one's), defined as
 $\hat{Q}_0= e^{-\pi \hat{P}_0 \hat{Z}_0/2}\hat{Z}_0 e^{\pi \hat{P}_0 \hat{Z}_0/2} = i\hat{Z}_0,  \hat{P}_0 = e^{-\pi \hat{P}_0 \hat{Z}_0/2}\hat{P}_0 e^{\pi \hat{P}_0 \hat{Z}_0/2} = -i\hat{\Pi}_0$.
 Hence the transformed Hamiltonian is given by
 \begin{equation}
 \hat{H}_{can} = i \hat{\Pi}_1 \hat{Z}_0 + \sum_{i=2}^{n} \hat{\Pi}_i \hat{Z}_{i-1} + \hat{\bar{\mc{H}}}.
 \label{imaginary-ham}
 \end{equation}
 The above equation is showing that all HD theories can be brought to this general form where antilinearity emerges out once the canonical Hamiltonian is defined in terms of the newly transformed Hamiltonian. This transformation \textbf{have} become a requirement for the HD theories due to the existence of momenta corresponding to the HD fields \cite{bender1}.

    \section{The Gallilean invariant Chern-Simons model}
    In this section we consider a specific model to show the efficacy of the discussions {of}  the earlier sections. We consider the Gallilean invariant Chern-Simon's model which is given by 
    \begin{equation}
    \mc{L} = \frac{1}{2}m\dot{x}_i^2 - k \epsilon_{ij}\dot{x}_i \ddot{x}_j.
    \end{equation}
    This is a  {nonrelativistic model in two spatial dimensions} and $k$ having physical dimension of $[M][T]$. Being a higher derivative model, we convert this into a first order Lagrangian owing to the transformations
    \begin{eqnarray}
    q_{1i}  &=& x_i, \\ 
    q_{2i} &=& \dot{x}_{i}.
    \end{eqnarray}
With these new variables the first order Lagrangian takes the form as
\begin{equation}
\mc{L} = \frac{1}{2}m q_{1i}^2 - k \epsilon_{ij}q_{2i} \dot{q}_{2j} + \lambda_i(\dot{q}_{1i} - q_{2i}  ).
\end{equation} 
Here $\lambda_i$ are the Lagrange's multiplier incorporated to account the constraints $ \dot{q}_{1i} - q_{2i}$ arising due to the redefinition of the fields. If $\{p_{1i}, p_{2i}, p_{\lambda_i} \}$ are the momenta corresponding to the fields $\{q_{1i}, q_{2i}, \lambda_{i} \}$ then we get a set of primary constraints given by 
\begin{eqnarray}
\Phi_{i} = p_{1i} - \lambda_{i} \approx 0, \\
\psi_{i} = p_{2i} - k\epsilon_{ij}q_{2j} \approx 0, \\
\Xi_{i} = p_{\lambda_{i}} \approx 0.
\end{eqnarray}
All these three primary constraints are second class in nature due to their  non-zero Poisson brackets which are given by 
\begin{eqnarray}
\{ \psi_i , \psi_j \} &=&  - 2 k \epsilon_{ij}, \\
\{ \Phi_{i} , \Xi_{j} \} &=& -\delta_{ij}.
\end{eqnarray}
{
The second class constraints are removed by setting them to be zero. This also replaces all the Poisson brackets in the theory to Dirac bracket defined as
\begin{equation}
\{ \xi_i , \xi_j \}_D = \{ \xi_i , \xi_j \} - \{\xi_i, \psi_m \} \Delta^{-1}_{mn} \{\psi_n, \xi_{j} \}
\end{equation}
 In the present case, the set of phasespace variables $\xi_i$ is $\{ q_{1i}, q_{2i}, p_{1i}, p_{2i} \}$ and \textbf{the Poission brackkets between the second class constraints is defined by } $\Delta_{mn} = \{\psi_m , \psi_n \}$. The non-zero dirac brackets are given by
 \begin{eqnarray}
 &&
 \{ q_{1i}, p_{1j} \}_D = \delta_{ij} \\ && \nonumber 
 \{ q_{2i}, q_{2j}\}_D = \frac{1}{2k}\epsilon_{ij} \\ && \nonumber 
 \{ q_{2i}, p_{2j}\}_D = \frac{3}{2} \delta_{ij}  \\ && \nonumber 
 \{ p_{2i}, p_{2j} \}_D = - \frac{k}{2} \epsilon_{ij} 
 \end{eqnarray}
}
Using the usual definition of the  canonical Hamiltonian we can write for the present model  
\begin{equation}
\mc{H}_{can} = -\frac{m}{2}q_{2i}^2 + \lambda_iq_{2i}.
\end{equation}
Since the system has constraints we should consider the total Hamiltonian instead, which is obtained by adding the primary constraints linearly to the canonical Hamiltonian as
\begin{equation}
{H}_T = \mc{H}_{can} + \Lambda_{1i}\Phi_i + \Lambda_{2i} \psi_{i} + \Lambda_{3i} \Xi_{i}.
\end{equation}
The Hamiltonian written thus involves undetermined Lagrange's multipliers $\{ \Lambda_{1i}, \Lambda_{2i}, \Lambda_{3i} \}$ and can be determined by considering the time evolution of the constraints. The time evolution of the constraints may give rise to secondary and tertiary constraints. We consider the brackets between the constraints and the total Hamiltonian and after equating them to zero, the following values of Lagrange's multipliers are obtained,

\begin{eqnarray}
\Lambda_{3i} &=& 0, \\
\Lambda_{2i} &=& \frac{1}{2k} (mq_{2j} + \lambda_{j})\epsilon_{ji}, \\
\Lambda_{1i} &=& -q_{2i}.
\end{eqnarray}

We can remove the second class constraints by setting them to zero and considering the Dirac brackets in place of Poisson brackets. Consequently, in this case, the total Hamiltonian become equal to the canonical Hamiltonian
\begin{equation}
H_T = \mc{H}_{can}.
\end{equation} 
{Upto this, we have discussed the classical views of this Gallilean invariant Chern-Simon's model. In the corresponding quantum version, we want to see the antilinear symmetry is present in the system.} For that, we should analyze the $\mc{PT}-$symmetries of this Hamiltonian under the changes of the space-time coordinates. The model is true $\mc{PT}$ symmetric which can be seen from
\begin{eqnarray}
\mc{P} \hat{H}_{T}\mc{P}^{-1} &=& \hat{H}_{T}^\dagger, \\ 
(\mc{PT}) \hat{H}_{T} (\mc{PT})^{-1} &=& \hat{H}_{T}.
\end{eqnarray}
As discussed in Sec II this Hamiltonian contains linear momenta terms and hence the transition to the quantum picture is not possible as the states will be unstable. Due to the higher derivative nature, for  removal of these linear fields, we may consider a similarity transformation in the form of change of variables $\hat{q}_{1i} = i\hat{z}_i, \hat{p}_{1i} = -i\hat{p}_{zi} $ as suggested in \cite{bender1998}  and obtain the Hamiltonian as 
\begin{equation}
\hat{\tilde{H}}_{T} = - \frac{m}{2}\hat{q}_{2i}^2  -i\hat{p}_{zi} \hat{q}_{2i}. \label{similarity_H}
\end{equation}
The total Hamiltonian in (\ref{similarity_H}) is clearly non-Hermitian in nature and $\mc{PT}$ symmetric. Being a non-Hermitian Hamiltonian, it does not necessarily mean to have {complex energy eigenvalues}. As discussed in (\ref{similarity_H}), owing to the $\mc{PT}$ symmetric nature, the model is having real energy eigenvalues. 
The existence of antilinearity in the model is confirmed since the Hamiltonian obeys $\mc{P} \hat{H}_{T}\mc{P}^{-1} = \hat{H}_{T}^\dagger$. Hence, as a requirement of (\ref{11},\ref{12}), for this present model, we can find the  corresponding intertwining operators which can be written as 
 \begin{eqnarray}
 \hat{V} &=& e^{-\hat{Q}}, \\
 \hat{A} &=& e^{\hat{Q}/2}.
 \end{eqnarray}
 These two operators are unitarily equivalent. 
 To check the antilinearity of the Hamiltonian,  we have considered 
  \begin{equation}
 \hat{Q} = \alpha \hat{p}_{zi} \hat{p}_{2i} + \beta \hat{z}_i \hat{q}_{2i}.
 \end{equation} 
 Now, we calculate the similarity transformation of the total Hamiltonian. This is done by calculating the transformation of the individual fields and replacing their values. After some algebraic calculations, we obtain
 \begin{eqnarray}
 \hat{A} \hat{H}_{T} \hat{A}^{-1} &=& e^{\hat{Q}/2} \hat{H}_{T} e^{-\hat{Q}/2} = \hat{\tilde{H}}_{T},
 \end{eqnarray}
where $\hat{\tilde{H}}_T$ is given by 
\begin{eqnarray}
\nonumber
\hat{\tilde{H}}_T &=& -\frac{m}{2}\Big( \hat{q}_{2i}^2 \cosh 2\sqrt{\alpha\beta}-\frac{\hat{p}_{zi}^2}{2}D^2 (\cosh 2\sqrt{\alpha\beta}-1) \Big)+\Big(\frac{\hat{q}_{2i}^2}{D} \sinh 2\sqrt{\alpha\beta} -\frac{\hat{p}_{zi}^2}{2} D\sinh 2\sqrt{\alpha\beta} \Big) \\ &&   -i  \hat{p}_{zi}\hat{q}_{2i} \Big(\frac{m}{2}D\sinh 2\sqrt{\alpha\beta}  + \cosh 2\sqrt{\alpha\beta}\Big) . \label{H_PT}
\end{eqnarray}

{Here we have taken $D = \sqrt{\alpha / \beta}$}. We can make this Hamiltonian free from the linear momenta terms, so that the states will be free from the Ostrogradski instability, owing to the condition as given below

\begin{equation}
\frac{m}{2}D\sinh 2\sqrt{\alpha\beta}  +\cosh 2\sqrt{\alpha\beta} = 0 .
\end{equation}

{The relation between $\alpha$ and $\beta$ can be found from the above equation. However, there remain arbitrariness in one of the variables either $\alpha$ or $\beta$. The solutions will differ from model to model depending on the mass term $m$.} The states $\ket{\psi}$ which correspond to the Hamiltonian $H_T$ are related to $\ket{\tilde{\psi}}$  as 
\begin{equation}
\ket{\tilde{\Psi}} = e^{-Q/2} \ket{\Psi}.
\end{equation} 
Where $\ket{\tilde{\psi}}$ are the states for the systems free from the ghosts. Thus being a higher derivative model and possessing antilinearity, we have successfully removed the Ostrogradski ghost from the system.    \section{Conclusion}
    
 The higher derivative theories have been very useful in gravity  \cite{Cano:2019ore}, cosmology \cite{Cuzinatto:2018vjt, Chialva:2014rla} Fractals \cite{ Becker:2019fhi} etc and used over the time by many authors despite the existence of the Ostrogradski instability. To remove this instability, various attempts were made but none of them were a complete success due to the conditions inherent while developing the theory. In the present paper, we have considered another aspect of this problem of Ostrogradski ghosts  using the antilinearity property of the HD theories. Due to this, we can employ the ideas developed earlier using $\mc{PT}$-symmetries and see if the Ostrogradski instability is curable or not.

      We have seen that, in the HD theories, the presence of the antilinear symmetry in the Hamiltonian was not clear until a proper similarity transformation was made. After this transformation, the HD Hamiltonian became non-Hermitian and the antilinear symmetry also emerged out. The operator connecting the antilinear symmetry of the Hamiltonian, in this case, was identified by comparing the form of the  $\mc{C}$-operator of the $\mc{PT}-$symmetric theories. To illustrate the efficacy of this approach, we have considered the Gallilean invariant Chern-Simon's model in 2+1 dimensions. The model contains higher derivative Chern-Simon's term which, upon quantization, shows a connection with the Noncommutative theory \cite{lukiersky}. The Ostrogradski instability was still prevailing in the Hamiltonian. On the contrary, here, the higher derivative Lagrangian was transformed into first order Lagrangian and \textbf{thus} the canonical Hamiltonian was obtained. After a similarity transformation, using the newly defined operator, of the non-Hermitian  Hamiltonian all the linear momenta terms vanished under suitable condition that must be obeyed. This condition put some restriction on the mass term $m$ of the particle.
    
    	Thus we have successfully removed the linear momenta (Ostrogradski ghost) term from the HD theory employing properties due to antilinearity. However, since this was a constrained system, the Hamiltonian was considered in the reduced phase space. This approach should be applicable to other higher derivative theories  also where the Ostrogradski ghost will plague the Hamiltonian.

	 
\end{document}